# Graphene for Emission of Terahertz Radiation

S. A. Mikhailov

Institute of Physics, University of Augsburg, Augsburg, D-86135 Germany

E-mail: sergey.mikhailov@physik.uni-augsburg.de

**Abstract** – Graphene is a two-dimensional crystal consisting of a monatomic layer of carbon atoms. Electrons and holes in graphene behave as quasi-relativistic particles with zero effective mass and large (as compared to semiconductors) Fermi velocity. These unique physical properties can be used for designing detectors, emitters and modulators of terahertz radiation. Here we discuss several ideas of using graphene based structures for emission of terahertz radiation.

## I. Introduction

Graphene is a two-dimensional (2D) crystal consisting of a single layer of carbon atoms arranged in a hexagonal lattice.[1] It was experimentally discovered in 2004,[2] and attracted much interest due to its unique mechanical, thermal, electrical and optical properties. In contrast to many conventional materials electrons and holes in graphene have not parabolic but linear energy dispersion

$$E_{\pm}(\boldsymbol{k}) = \pm \hbar v_F |\boldsymbol{k}| \qquad (1)$$

with the very high, as compared to semiconductors, Fermi velocity $v_F \approx 10^8$ cm/s. Graphene is a semimetal, with the typical charge carrier density $\sim 10^{11} - 10^{13}$ cm$^{-2}$, and its plasma frequency lies in the terahertz (THz) range. Together with other 2D crystals[3] such as dielectric boron nitride (BN) or semiconducting transition metal dichalcogenides, it may serve as a basis for THz electronic devices, such as emitters, detectors and modulators. Here we discuss some ideas of using graphene and graphene related materials for emission of THz radiation.

## II. Smith-Purcell-type terahertz emitter

When a fast electron beam moves, with the average velocity $v$, in a periodic potential with the period $a$, the momenta and velocities of electrons oscillate and they emit electromagnetic waves with the frequency $f = v/a$ (the Smith-Purcell effect). This physical principle is used in vacuum devices such as the backward wave oscillators or free electron lasers. The same effect was expected to work in semiconductor quantum-well structures, in which 2D electrons are driven with a large velocity under the gate made in the form of a periodic grating. However experiments[4] showed that, instead of the Smith-Purcell strong coherent radiation, the system emits weak thermal radiation at the frequency of 2D plasmons $f_p$. In Ref. 5 we showed that the Smith-Purcell formula is valid only under the condition $f_p \ll v/a$ which was not satisfied in Ref. 4. If the plasma frequency of 2D electrons is not so small, which is typical for semiconductors quantum well structures, the Smith-Purcell-type emission should be observed at the frequency $= v/a - f_p$, and only if the electron velocity exceeds the threshold value

$$v > v_{th} \approx f_p a = \sqrt{\frac{n_s e^2 a}{m \epsilon}}, \qquad (2)$$

where $n_s$ and $m$ are the density and effective mass of 2D electrons and $\epsilon$ is the dielectric constant of surrounding medium. In semiconductor 2D electron gas systems the condition (2) is very difficult to satisfy, if possible at all.

In Ref. 6 we have shown that graphene may help to overcome this difficulty. Indeed, the Fermi velocity in graphene is about one order of magnitude larger than in semiconductor structures. In addition, due to the vanishing effective mass of graphene electrons, they can be accelerated up to the large drift velocity by a much smaller dc electric field, Fig. 1.

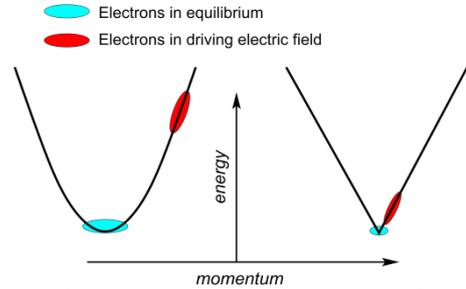

Fig. 1. In order to accelerate a small amount of electrons up to velocities of order of $10^8$ cm/s one needs a large change of the momentum in the case of the parabolic energy dispersion (semiconductors, left) and a much smaller change of the momentum in the case of the linear energy dispersion (graphene, right).

However, a continuous graphene layer is not suitable for the realization of the Smith-Purcell emitter because of two reasons. First, according to (2) the density of electrons should be very small (smaller than $10^{10}$ cm$^{-2}$ according to estimates) while in continuous graphene it is usually substantially larger. Second, due to the specific energy dispersion of graphene electrons (1), the oscillations of momenta in graphene do not result in the oscillations of their velocities and hence of the current. The both problems can be solved if to use, instead of the continuous graphene layer, an array of narrow graphene stripes with the width $W$. Then a gap $\Delta \sim 1/W$ opens in the spectrum, the electron velocity becomes momentum dependent, and the density of electrons can be made small if the chemical potential lies in the gap. A detailed analysis of the graphene-based Smith-Purcell-type THz emitter[6,7] showed that the threshold conditions of radiation (2) can be satisfied indeed, and the proposed device can emit radiation in the frequency range ~0.1–30 THz with the power density up to ~0.5 W/cm$^2$ at room temperature. The proposed in Refs. 6,7 structure consists of an array of graphene stripes lying on a substrate and separated from the grating gate by a thin (few nm) dielectric BN layer. The grating is assumed to be made out of another graphene layer having the form of narrow stripes oriented perpendicular to those of the first layer. A dc bias is assumed to be applied between the gate and the first graphene layer which leads to a periodic potential $U(x) = U(x - a)$ seen by graphene electrons of the first layer. The dielectric layer separating the main conducting layer and the grating can be much thinner than in semiconductor quantum-well heterostructures which enhances the interaction of the two plasmas. Figure 2 shows the time dependence of the electron velocity in the proposed emitter calculated[6] with experimentally realizable parameters. The non-sinusoidal shape of the current response presented by the red curve is due to the step-like form of the periodic potential $U(x)$ and the graphene-specific nonlinear velocity-momentum relation.

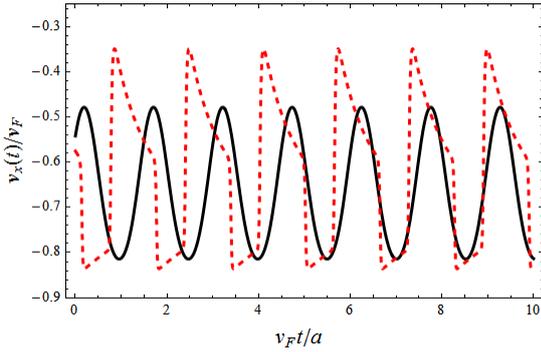

Fig. 2. The time dependence of the velocity of electrons in the proposed graphene – BN THz emitter at a set of typical, experimentally realizable parameters, for details see Ref. 6. Two curves differ by the model of the periodic potential created by the dc voltage applied to the grating-gate: black and red curves correspond to a sinusoidal and a step-like potentials $U(x)$, respectively.

An important parameter of the graphene-BN based THz emitter is the ratio of the mean free path $l$ to the grating period which should be large. In Ref. 6 we assumed $l$ to be 1 μm. Later the experimentally achieved mean free path values exceeding 28 μm have been reported in CVD grown graphene fully encapsulated in hexagonal BN.[8] Availability of such a high-quality graphene opens a direct path towards the realization of the proposed[6,7] graphene-BN-based tunable and powerful THz emitters.

### III. NONLINEAR ELECTRODYNAMIC RESPONSE OF GRAPHENE

Another opportunity to use graphene in THz electronics is related to the fact that it demonstrates a strongly nonlinear electrodynamic response. It was predicted[9] that the linear energy dispersion (1) should lead to a strongly nonlinear electrodynamic response of this material. This can be seen from simple physical considerations. In contrast to massive particles, the velocity $\boldsymbol{v}_\pm = \partial E_\pm(\boldsymbol{k})/\partial \boldsymbol{p} = \pm v_F \boldsymbol{k}/|\boldsymbol{k}|$ of electrons with the spectrum (1) is not proportional to the momentum, therefore if an electric field $\boldsymbol{E}_0 \cos(\omega t)$ acts on such electron, its momentum $\boldsymbol{p} = \hbar \boldsymbol{k}$ oscillates as $\sin(\omega t)$ but the velocity $v \sim \text{sgn}[\sin(\omega t)]$, and hence the current, contain higher frequency harmonics. All other nonlinear effects can evidently also be seen in graphene due to the same reason.

The qualitative picture of the nonlinear graphene response outlined above is valid at low (microwave, THz) frequencies $\hbar\omega < 2E_F$ ($E_F$ is the Fermi energy) when the inter-band transitions between the valence and conduction bands in graphene can be neglected. A quantum theory of the third-order nonlinear response of graphene which takes into account both the intra- and inter-band transitions and is valid at all frequencies was developed in Refs. 10,11. Figure 3 shows an example of the calculated efficiency of the four-wave mixing (FWM) process $\omega_1, \omega_2 \to 2\omega_1 - \omega_2$ with the output frequency lying in the THz band. Shown is the efficiency of the FWM frequency transformation

$$\eta^{(3)} = \frac{I_{2\omega_1 - \omega_2}}{I_{\omega_1}^2 I_{\omega_2}} \qquad (3)$$

as a function of the Fermi energy at the input frequencies 30 and 59 THz and the output frequency 1 THz ($I$ is the intensity of radiation). The sharp resonant features at $T = 0$ are related to the inter-band transition at the Fermi edge.

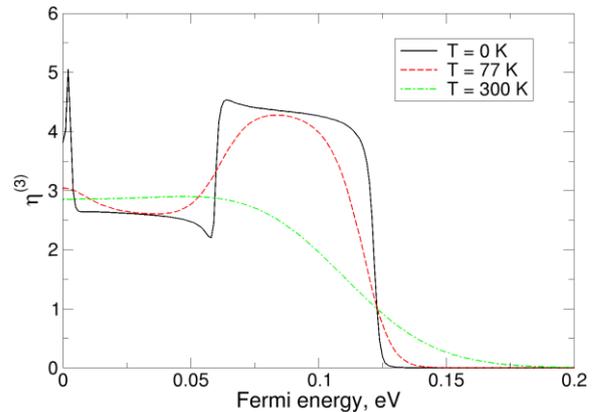

Fig. 3. The efficiency (3) of the four-wave mixing process $\omega_1, \omega_2 \to 2\omega_1 - \omega_2$, with the input frequencies $f_1 = \omega_1/2\pi = 30$ THz and $f_2 = 59$ THz and the output frequency 1 THz, as a function of Fermi energy. The efficiency $\eta^{(3)}$ is shown in units $10^{-19}$ (cm$^2$/W)$^2$.

Other opportunities of the THz wave emission are related to the third harmonic generation (THG) in graphene. The efficiency of this process in the isolated graphene was calculated, e.g., in Ref. 10. In Refs. 12,13 it was shown that the substrate which support the graphene layer may substantially influence the THG efficiency. In particular, if graphene lies on a dielectric substrate with a metalized back side, the third harmonic intensity can be increased by two orders of magnitude as compared to the isolated graphene.[12,13] Placing graphene on a substrate made out of a polar dielectric may lead to another useful effect. The THG efficiency in the isolated graphene falls down very quickly with the fundamental harmonic frequency ω, as $1/\omega^3$ at low ($\hbar\omega < 2E_F$) and as $1/\omega^4$ at high ($\hbar\omega > 2E_F$) frequencies. If graphene lies on the polar dielectric substrate the THG efficiency becomes practically frequency independent if the input wave frequency ω lies in the reststrahlen band between the transverse and longitudinal optical phonon frequencies, $\omega_{TO} < \omega < \omega_{LO}$.[13] In addition, in this frequency range the THG efficiency is also larger than in the isolated graphene layer in air.

A very useful and unique feature of the nonlinear graphene response is that the efficiency of different nonlinear processes can be electrically controlled by the gate voltage since the third-order response functions of graphene have strong resonances at frequencies corresponding to the inter-band transitions at the Fermi edge. For the first time it was demonstrated in the FWM process in Ref. 14.

The third-order processes considered above assumes the local ($\boldsymbol{q} = 0$) nonlinear response of graphene,[10,11] where $\boldsymbol{q}$ is the wave-vector of the input electromagnetic wave parallel to the 2D layer. If $\boldsymbol{q} \neq 0$, second-order nonlinear processes which can also be used for THz emission take place in graphene. One of the second-order nonlinear phenomena, the nonlinear difference frequency generation, was theoretically studied in Ref. 15. At the finite wave-vectors $\boldsymbol{q}$ one can additionally use the advantage of the 2D plasmon excitation in graphene and the resonance enhancement of the nonlinear effects[16] at the 2D plasmon frequencies.

## IV. SUMMARY AND CONCLUSIONS

As a consequence of its unique electronic properties graphene demonstrates unusual and very useful electrodynamic properties, in particular, the nonlinear electrodynamic and optical response. These properties can be used for the creation of different optoelectronic and photonic devices at THz, infrared and optical frequencies. We have briefly overviewed a small part of possible graphene applications related to the emission of THz radiation. Detectors and modulators of THz radiation have also been proposed and experimentally studied in the literature. All this makes graphene a very attractive material for THz electronics.

## ACKNOWLEDGMENTS

This work was funded by the European Union's Horizon 2020 research and innovation programme Graphene Core 2 under Grant Agreement No. 785219.